\documentclass[prl,twocolumn,twoside,preprintnumbers,superscriptaddress,nofootinbib,showpacs]{revtex4-2}
\usepackage[colorlinks=true,citecolor=blue,filecolor=blue,linkcolor=blue,urlcolor=blue,pdftex]{hyperref}

\makeatletter
\let\latex@fnsymbol\@fnsymbol
\renewcommand\@fnsymbol[1]{\ensuremath{\ifcase#1\or * \or \dagger\or \ddagger\or
   \mathsection\or \mathparagraph\or \|\or **\or \dagger\dagger
   \or \ddagger\ddagger\or ***\or **** \else\@ctrerr\fi}}
\makeatother

\usepackage{xspace}
\usepackage[usenames,dvipsnames]{color}
\usepackage{booktabs,graphicx,mathrsfs,verbatim,amsmath,units,soul}
\usepackage{xspace} 
\usepackage{upgreek}
\usepackage{textcomp}
\usepackage{gensymb}
\usepackage{tabularx}
\usepackage{xcolor}
\usepackage{multirow}
\usepackage{tabularx}
\usepackage[separate-uncertainty=true]{siunitx}
\DeclareSIUnit\clight{\text{\ensuremath{c}}}
\DeclareSIUnit\evm{\eV\per\clight\squared}
\DeclareSIUnit\year{\text{y}}
\DeclareSIUnit\tonneyears{\tonne\times\year}
\DeclareSIUnit\evm{\eV\per\clight\squared}
\DeclareSIUnit\gevm{\GeV\per\clight\squared}
\DeclareSIUnit\PE{\text{PE}}
\DeclareSIUnit\ph{\text{ph}}
\DeclareSIUnit\pperkev{\ph\per\kev}
\DeclareSIUnit\events{\text{events}}
\DeclareSIUnit\pertonneyears{\per(\tonne\times\year)}
\DeclareSIUnit\mev{\mega\eV}

\newcommand{\ton}{ton}

\newcommand{\Stwo}{\ensuremath{\mathrm{S2}}\xspace}

\newcommand{\qy}{\ensuremath{\mathrm{Q}_\mathrm{y}}\xspace}
\newcommand{\ly}{\ensuremath{\mathrm{L_\mathrm{y}}}\xspace}

\newcommand{\F}{\ensuremath{\Phi\xspace}}
\newcommand{\sigmac}{\ensuremath{\sigma_\cevns}}

\newcommand{\gevcsq}{\ensuremath{\mathrm{GeV}/c^2}}
\newcommand{\gevm}{\gevcsq}
\newcommand{\prevstwodt}{\ensuremath{\Stwo_\mathrm{prev}/\Delta t_\mathrm{prev}}}
\newcommand{\kev}{\ensuremath{\mathrm{keV}}\xspace}

\newcommand{\cevns}{{\text{CE}\ensuremath{\nu}\text{NS}}\xspace}
\newcommand{\cevnsnospace}{{\text{CE}\ensuremath{\nu}\text{NS}}}

\newcommand{\Z}{\ensuremath{Z\xspace}}

\newcommand{\beight}{\ensuremath{{}^8\mathrm{B} }\xspace}



\newcommand{\lagr}{\mathcal{L}} 
\newcommand{\xesub}{\ensuremath{\mathrm{Xe1T}}}

\newcommand{\nuis}{\theta}
\newcommand{\nuiss}{\vec{\nuis}}

\newcommand{\GBDT}{\text{GBDT}\xspace}
\newcommand{\itsec}[1]{{\it #1 }--- }
\newcommand{\fref}[1]{Fig.~\ref{#1}}
\newcommand{\tref}[1]{Tab.~\ref{#1}}
\newcommand{\eref}[1]{Eq.~\ref{#1}}

\begin{document}
\newcounter{includeappendix}
\setcounter{includeappendix}{1} 

\title{Search for Coherent Elastic Scattering of Solar~\beight Neutrinos in the XENON1T Dark Matter Experiment}


\newcommand{\bologna}{\affiliation{Department of Physics and Astronomy, University of Bologna and INFN-Bologna, 40126 Bologna, Italy}}
\newcommand{\chicago}{\affiliation{Department of Physics \& Kavli Institute for Cosmological Physics, University of Chicago, Chicago, IL 60637, USA}}
\newcommand{\coimbra}{\affiliation{LIBPhys, Department of Physics, University of Coimbra, 3004-516 Coimbra, Portugal}}
\newcommand{\columbia}{\affiliation{Physics Department, Columbia University, New York, NY 10027, USA}}
\newcommand{\lngs}{\affiliation{INFN-Laboratori Nazionali del Gran Sasso and Gran Sasso Science Institute, 67100 L'Aquila, Italy}}
\newcommand{\mainz}{\affiliation{Institut f\"ur Physik \& Exzellenzcluster PRISMA, Johannes Gutenberg-Universit\"at Mainz, 55099 Mainz, Germany}}
\newcommand{\heidelberg}{\affiliation{Max-Planck-Institut f\"ur Kernphysik, 69117 Heidelberg, Germany}}
\newcommand{\munster}{\affiliation{Institut f\"ur Kernphysik, Westf\"alische Wilhelms-Universit\"at M\"unster, 48149 M\"unster, Germany}}
\newcommand{\nikhef}{\affiliation{Nikhef and the University of Amsterdam, Science Park, 1098XG Amsterdam, Netherlands}}
\newcommand{\nyuad}{\affiliation{New York University Abu Dhabi, Abu Dhabi, United Arab Emirates}}
\newcommand{\purdue}{\affiliation{Department of Physics and Astronomy, Purdue University, West Lafayette, IN 47907, USA}}
\newcommand{\rpi}{\affiliation{Department of Physics, Applied Physics and Astronomy, Rensselaer Polytechnic Institute, Troy, NY 12180, USA}}
\newcommand{\rice}{\affiliation{Department of Physics and Astronomy, Rice University, Houston, TX 77005, USA}}
\newcommand{\stockholm}{\affiliation{Oskar Klein Centre, Department of Physics, Stockholm University, AlbaNova, Stockholm SE-10691, Sweden}}
\newcommand{\subatech}{\affiliation{SUBATECH, IMT Atlantique, CNRS/IN2P3, Universit\'e de Nantes, Nantes 44307, France}}
\newcommand{\torino}{\affiliation{INAF-Astrophysical Observatory of Torino, Department of Physics, University  of  Torino and  INFN-Torino,  10125  Torino,  Italy}}
\newcommand{\ucsd}{\affiliation{Department of Physics, University of California San Diego, La Jolla, CA 92093, USA}}
\newcommand{\wis}{\affiliation{Department of Particle Physics and Astrophysics, Weizmann Institute of Science, Rehovot 7610001, Israel}}
\newcommand{\zurich}{\affiliation{Physik-Institut, University of Z\"urich, 8057  Z\"urich, Switzerland}}
\newcommand{\paris}{\affiliation{LPNHE, Sorbonne Universit\'{e}, Universit\'{e} de Paris, CNRS/IN2P3, Paris, France}}
\newcommand{\freiburg}{\affiliation{Physikalisches Institut, Universit\"at Freiburg, 79104 Freiburg, Germany}}
\newcommand{\lal}{\affiliation{Universit\'{e} Paris-Saclay, CNRS/IN2P3, IJCLab, 91405 Orsay, France}}
\newcommand{\napels}{\affiliation{Department of Physics ``Ettore Pancini'', University of Napoli and INFN-Napoli, 80126 Napoli, Italy}}
\newcommand{\nagoya}{\affiliation{Kobayashi-Maskawa Institute for the Origin of Particles and the Universe, and Institute for Space-Earth Environmental Research, Nagoya University, Furo-cho, Chikusa-ku, Nagoya, Aichi 464-8602, Japan}}
\newcommand{\laquila}{\affiliation{Department of Physics and Chemistry, University of L'Aquila, 67100 L'Aquila, Italy}}
\newcommand{\tokyo}{\affiliation{Kamioka Observatory, Institute for Cosmic Ray Research, and Kavli Institute for the Physics and Mathematics of the Universe (WPI), the University of Tokyo, Higashi-Mozumi, Kamioka, Hida, Gifu 506-1205, Japan}}
\newcommand{\kobe}{\affiliation{Department of Physics, Kobe University, Kobe, Hyogo 657-8501, Japan}}
\newcommand{\ucla}{\affiliation{Physics \& Astronomy Department, University of California, Los Angeles, CA 90095, USA}}
\newcommand{\kit}{\affiliation{Institute for Astroparticle Physics, Karlsruhe Institute of Technology, 76021 Karlsruhe, Germany}}
\newcommand{\tsinghua}{\affiliation{Department of Physics \& Center for High Energy Physics, Tsinghua University, Beijing 100084, China}}
\newcommand{\alsoatferrara}{\affiliation{INFN, Sez. di Ferrara and Dip. di Fisica e Scienze della Terra, Universit\`a di Ferrara, via G. Saragat 1, Edificio C, I-44122 Ferrara (FE), Italy}}
\newcommand{\alsoatsuny}{\affiliation{Simons Center for Geometry and Physics and C. N. Yang Institute for Theoretical Physics, SUNY, Stony Brook, NY, USA}}
\newcommand{\alsoatutrecht}{\affiliation{Institute for Subatomic Physics, Utrecht University, Utrecht, Netherlands}}
\newcommand{\alsoatcoimbrapoli}{\affiliation{Coimbra Polytechnic - ISEC, Coimbra, Portugal}}
\newcommand{\alsoatiarnagoya}{\affiliation{Institute for Advanced Research, Nagoya University, Nagoya, Aichi 464-8601, Japan}}



\author{E.~Aprile}\columbia
\author{J.~Aalbers}\stockholm
\author{F.~Agostini}\bologna
\author{S.~Ahmed Maouloud}\paris
\author{M.~Alfonsi}\mainz
\author{L.~Althueser}\munster
\author{F.~D.~Amaro}\coimbra
\author{S.~Andaloro}\rice
\author{V.~C.~Antochi}\stockholm
\author{E.~Angelino}\torino
% ...

\collaboration{XENON Collaboration}\email[]{xenon@lngs.infn.it}
\noaffiliation

\date{\today}

\noaffiliation

\date{\today}   

\begin{abstract}
We report on a search for nuclear recoil signals from solar \beight neutrinos elastically scattering off xenon nuclei in XENON1T data, lowering the energy threshold from \SI{2.6}{\kev} to \SI{1.6}{\kev}.
We develop a variety of novel techniques to limit the resulting increase in backgrounds near the threshold.
No significant \beight neutrino-like excess is found in an exposure of \SI{0.6}{\tonneyears}. 
For the first time, we use the non-detection of solar neutrinos to constrain the light yield from 1-2 keV nuclear recoils in liquid xenon, as well as non-standard neutrino-quark interactions. 
Finally, we improve upon world-leading constraints on dark matter-nucleus interactions for dark matter masses between \SI{3}{\gevm} and \SI{11}{\gevm} by as much as an order of magnitude. 
\end{abstract}

\maketitle

\itsec{Introduction}
Neutrinos from the Sun, atmospheric cosmic-ray showers, and supernovae can produce observable nuclear recoils (NRs) via coherent elastic scattering off nuclei in liquid xenon (LXe) detectors searching for dark matter (DM)~\cite{billard}.
The coherent elastic neutrino-nucleus scattering (\cevnsnospace) process~\cite{cevnsproposal,COHERENTdiscovery,conus,connie}
produces the same signature as the one expected from DM-nucleus interactions, and thus the two can only be distinguished by their recoil spectra.
Solar \beight neutrinos are expected to contribute the greatest number of \cevns events in LXe DM search experiments.
These events fall near the energy thresholds of such detectors, with a spectrum indistinguishable from \SI{6}{\gevm}~DM.

The XENON1T dark matter search experiment, operated at the INFN Laboratori Nazionali del Gran Sasso (LNGS) until Dec. 2018, used a sensitive target of \SI{2.0}{\tonne} of LXe in a two-phase time projection chamber (TPC).
Two arrays of photomultiplier tubes (PMTs) at the top and bottom of the TPC allowed simultaneous detection of scintillation light (S1) and, via electroluminescence, ionization electrons (S2)~\cite{xenon1t_instrument, xenon1t_pmt}. 
With the largest exposure of any LXe TPC, data from XENON1T has been used to search for a variety of DM candidates, resulting in world-leading upper limits for DM-nucleus interactions~\cite{xenon1t_sr1,xenon1t_s2only,xenon1t_SD,xenon1t_pion}. 
Though no excess of \cevns from \beight neutrinos (\beight~\cevns) was observed due to the energy threshold in these analyses, 
they will soon become an important background given the large exposures of next-generation multi-\ton~LXe detectors~\cite{pandax_sensitivity,lz_sensitivity, xenonnt_sensitivity}.
In this Letter, we present a search for \beight~\cevns events in XENON1T data between Feb.~2, 2017 and Feb.~8, 2018 (``SR1" in Ref.~\cite{xenon1t_sr1}). In this new analysis, we achieve unprecedented sensitivity by reducing the energy threshold. 

\itsec{Analysis Strategy} 
The \beight~\cevns expectation in XENON1T depends on: the \beight neutrino flux \F, measured~\cite{borexino_b8, SNO} as \SI{5.25 \pm 0.20e6}{\per\cm\squared\per\second}; the \cevns cross section, from the Standard Model; the nuclear recoil scintillation light yield in xenon \ly; and the ionization yield \qy.
We first present a search for \beight~\cevns events in XENON1T, expecting $2.1$ \cevns events given nominal estimates of the above variables.
We then combine XENON1T data with external measurements, as appropriate, to constrain these variables.
We constrain \ly by considering external measurements of \qy and \F. 
Next, by including external measurements of \qy and \ly, we use XENON1T data to determine \F~independently.
We also constrain non-standard neutrino interactions by relaxing the standard model assumption on the \cevns~cross section. 
Finally, by considering \beight~\cevns as a background and applying external constraints on all variables, we use the data to set limits on DM-nucleus interactions.

\begin{figure}[ht]
\includegraphics[width=1.\columnwidth]{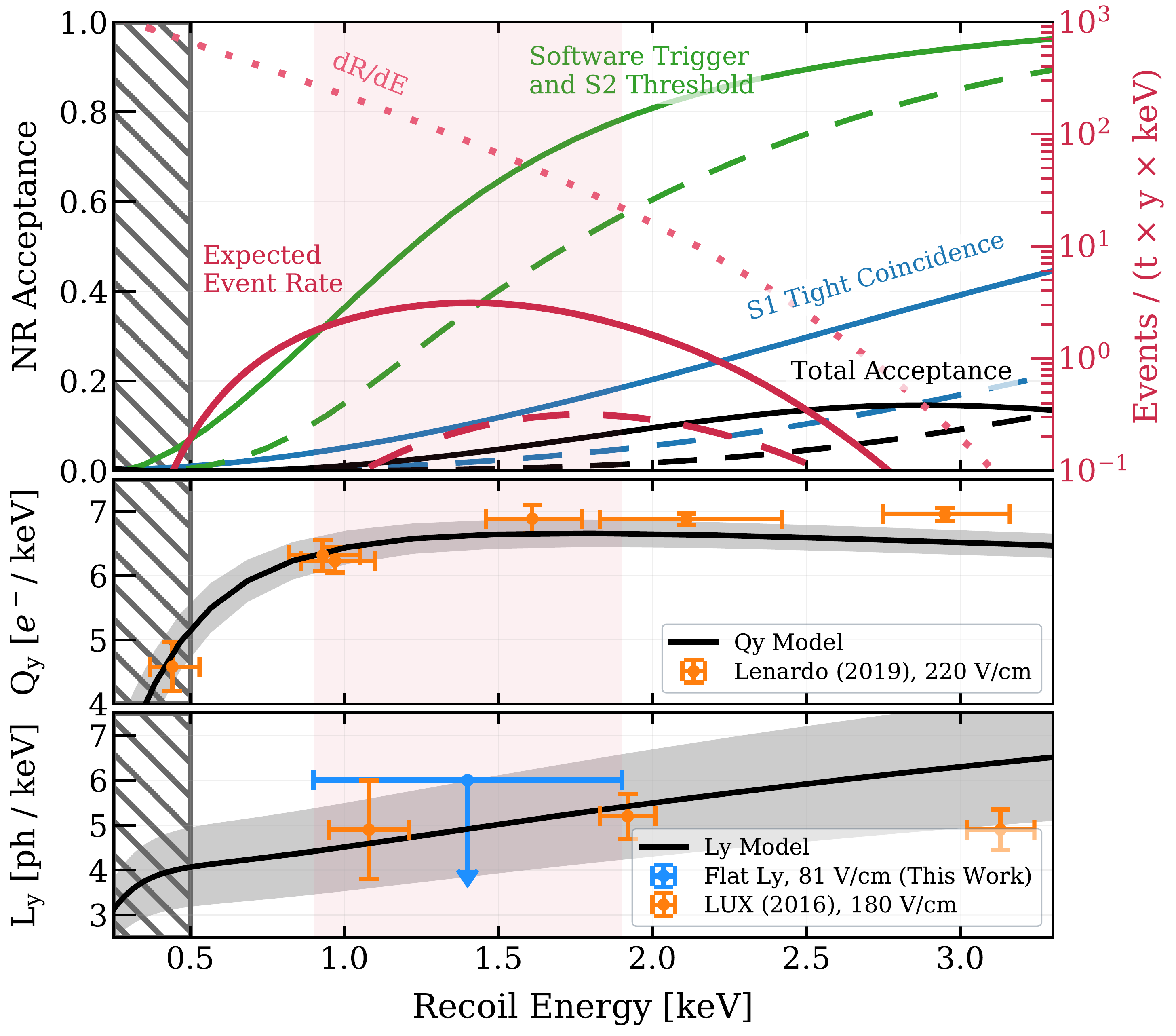}
\caption{
Top: Improvement of the NR acceptance in this work (solid) with respect to previous DM analyses (dashed)~\cite{xenon1t_sr0,xenon1t_sr1}, including S1 detection efficiency (blue), software trigger and S2 threshold acceptance (green), and total acceptance after other quality and background rejection cuts (black). The right axis shows the recoil spectrum of \beight~\cevns or dark matter of mass \SI{6}{\gevm} and cross section \SI{4e-45}{\cm^2} (dotted pink), and the products of this spectrum with the total acceptances (red) as a function of true recoil energy. The acceptances and resulting spectra are based on the nominal (NEST) yield models. The red shaded interval contains \SI{68}{\percent} of expected CEvNS events. 
Middle: The most precise available measurements of \qy \cite{lenardo} (orange), with the \qy model described in the text overlaid (black).
Bottom: Constraints on \ly (in photons per \kev) from LUX (orange)~\cite{lux_dd}, and the \SI{68}{\percent}~upper limit from this work described in the Results section (blue), with the \ly model described in the text overlaid (black). To be conservative, no response is assumed below the \SI{0.5}{\kev} cutoff (hatched gray).}\label{fig:efficiency_and_spectrum}
\end{figure}

\itsec{\cevns signal} 
The expected recoil spectrum of \beight~\cevns in LXe is shown in~\fref{fig:efficiency_and_spectrum}~(top, dotted red). 
The scintillation and ionization responses are relatively uncertain at \beight\,\cevns energies ($<\SI{2}{\kev}$), and NR calibration measurements in XENON1T scarcely overlap this region, instead producing S1s and S2s similar to DM of mass $\geq\SI{30}{\gevm}$.
Therefore, we modify the NR model in~\cite{xenon1t_sr1,xenon1t_analysis2} by decoupling the light and charge yields to allow for additional freedom. 

The NR charge yield \qy has been measured down to \SI{0.3}{\kev}~\cite{lenardo}, providing strong constraints at \beight\,\cevns energies which are included in v2.1.0 of the NEST package~\cite{nest}.
We use the best fit and uncertainty from NEST to define the shape of \qy, fitting a single free ``interpolation parameter'' $q$ to the measurements which specifies \qy within this uncertainty, resulting in the model shown in~\fref{fig:efficiency_and_spectrum}~(middle). The central black line (edges of the shaded interval) in the figure corresponds to $q$ equaling $0$~($\pm1$).
Measurements of the LXe NR light yield \ly~\cite{lux_dd} have a large ($\approx\SI{20}{\percent}$) uncertainty near \SI{1}{\kev}. 
Since the NEST \ly uncertainty is largely set by measurements at energies far above our region of interest (ROI),
we fit these measurements using a free parameter that scales the NEST best fit \ly.
These measurement and the resulting model are shown in~\fref{fig:efficiency_and_spectrum}~(bottom).
The \ly and \qy parameter fits use external measurements between $0.9$ and \SI{1.9}{\kev}, a central interval containing \SI{68}{\percent} of expected \beight~\cevns events after all acceptance losses.
We conservatively assume zero \ly below~\SI{0.5}{\kev}, the lowest energy measurement available~\cite{dongqing_thesis}.
This treatment has a percent-level effect on the expected~\cevns~rate, since the detection efficiency below this ``cutoff energy'' is $< \num{e-3}$.

The XENON1T S1 detection threshold was previously limited by the requirement 
that three or more PMTs detect pulses above threshold (denoted as ``hits'')
within \SI{50}{\nano\second}~\cite{xenon1t_analysis1}, leading to a 1\% acceptance of \cevns recoils above the \SI{0.5}{\keV} cutoff.
We reduce this ``tight-coincidence'' requirement to two hits within \SI{50}{\nano\second}, increasing the total acceptance above the \SI{0.5}{\keV} cutoff to \SI{5}{\percent}. 
Another efficiency loss comes from \beight~\cevns~S2s failing the software trigger, which requires 60 significant PMT signals~\cite{xenon1t_daq}, or the S2 analysis threshold.
The sensitivity is therefore impaired by the presence of electronegative impurities in the LXe, which reduce S2s along the drift path. 
The 120 PE S2 analysis threshold, reduced from 200 PE, accepts \SI{92}{\percent} of \cevns events that pass the software trigger.
Acceptance losses due to new event selection criteria introduced to ress backgrounds are described below. 
\fref{fig:efficiency_and_spectrum} (top) shows the S1 tight-coincidence acceptances, software trigger and S2 threshold acceptances, and total acceptances for this and previous analyses, and the resulting spectra of expected \beight~\cevns events. 
The Supplemental Material of this Letter provides details on the waveform simulation used to calculate all acceptances, and demonstrates excellent matching between real and simulated S1s and S2s. The overall change in acceptance results in a lowering of the energy threshold, defined as the energy where \SI{5}{\percent} of recoils are detected, from \SI{2.6}{\kev} to \SI{1.6}{\kev}.
The ROI for the \cevns search is defined by S2s between 120 and 500~ photoelectrons (PE), and S1s between 1.0 and \SI{6.0}{\PE} consisting of two or three hits. 
In this ROI, the \beight~\cevns signal expectation increases twentyfold with respect to previous NR searches~\cite{xenon1t_sr1, xenon1t_SD, xenon1t_pion} because of the relaxed tight-coincidence requirement and lower S2 threshold, derived from integrating the expected event rate in \fref{fig:efficiency_and_spectrum} (top). 
Because of the minimal overlap with previously studied data, we consider this a blind analysis.

\itsec{Backgrounds}
This analysis considers all backgrounds described in~\cite{xenon1t_sr1,xenon1t_analysis2}.
Radon daughters decaying on the inner surface of the TPC wall produce events with reduced S2s, contributing to the background in the ROI. 
In order to reduce this background to a negligible level, we use a fiducial volume of \SI{1.04}{\tonne}, similar to the one chosen for~\cite{xenon1t_sr0} but smaller than the one used in~\cite{xenon1t_sr1}. 

The accidental coincidence (AC) of S1 and S2 peaks incorrectly paired by the XENON1T reconstruction software mimics real interactions. 
AC background events are modeled by sampling (with replacement) from isolated S1s and S2s and assigning a random time separation between them.
Most S1s contributing to AC events originate from the pileup of lone hits from individual PMTs. Other sources include low-energy events occurring below the cathode or on the inner detector surface, and light leaking inside the active volume. 
AC forms the dominant background for this search, since the overall rate of isolated S1s increases by two orders of magnitude when we require only two hits. 
The rate and distribution of isolated S1s are determined using S1 peaks found in the extended event window of \SI{1}{m\s} before the S1 of high-energy events, as in \cite{xenon1t_sr1,xenon1t_analysis2}. 
For this analysis, the data is reprocessed with an updated algorithm~\cite{pax} to better retain the isolated S1s preceding these high-energy events, eliminating the dominant systematic uncertainty in the AC rate~\cite{xenon1t_sr1}.

High-energy events from gamma-ray backgrounds can also contaminate subsequent events with lone hits, a dominant source of S1s in this analysis.
For each event, the preceding event with the highest potential to produce lone hits is identified by dividing its largest S2 area by its time difference from the current event, denoted as \prevstwodt.
The selection \prevstwodt$<\SI{12}{\PE\per\micro\second}$ reduces the rate of isolated S1s
by \SI{65}{\percent}, accepting \SI{87}{\percent} of \beight~\cevns signals. 
Furthermore, we require the PMT signal sum within the first \SI{1}{m\s} of an event to be $<\SI{40}{\PE}$ and that this interval contains at most a single S1, accepting \SI{96}{\percent} of remaining events. 
After these selections, the total isolated-S1 
rate is \SI{11.2}{\hertz}, ten times higher than for a threefold tight-coincidence requirement~\cite{xenon1t_sr1}. 
The total exposure after these selection criteria is \SI{0.6}{\tonneyears}.

The same high-energy events can also produce small S2s appearing in subsequent events \cite{lux_electronbg}, potentially leading to unaccounted-for correlations between the isolated-S1 and isolated-S2 samples. 
In order to reduce these correlations, we further require that no S2 signal is found within the first millisecond of the event, and apply a cut on the horizontal spatial distance between the current and previous S2. 
These selections, together with the selection on \prevstwodt, allow us to model the AC background for S2s down to~\SI{80}{\PE} and reduce the isolated-S2 event rate therein to $\SI{1.0}{\milli\hertz}$. For comparison, the isolated-S2 event rate in~\cite{xenon1t_sr1} was $\SI{2.6}{\milli\hertz}$ for S2s above~\SI{100}{\PE}~\cite{xenon1t_sr1}.

\begin{figure}[t]
    \centering
    \includegraphics[width=0.9\columnwidth]{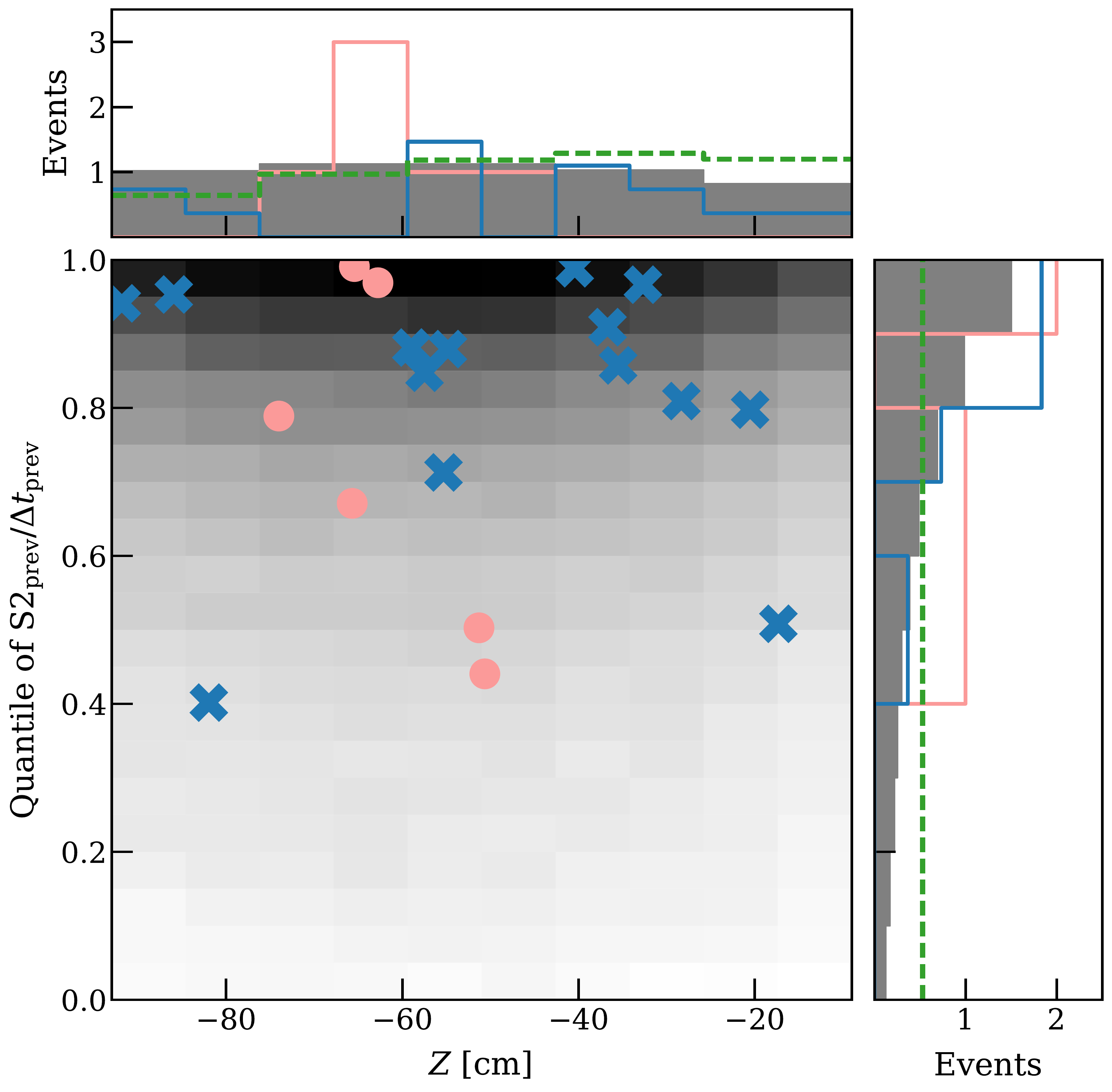}
    \caption{Events in the science dataset (pink circles) and the AC-enriched validation region (blue crosses) projected onto \Z~and the quantile of the \prevstwodt~value for NR signals. The AC model is shown in gray. Smaller panels show the projection of the model and data onto each axis, as well as the \beight~\cevns model (green dashed), normalized to its upper limit. The AC-enriched region data in blue has a slightly different \Z distribution due to the inverted GBDT cut, but is included for illustration, scaled by $0.36$, the ratio of expected AC events in each dataset.}
    \label{fig:datascatter}
\end{figure}

Selections that require both S1 and S2, such as the fiducial volume and S2 signal width~\cite{xenon1t_analysis1} cuts (which depend on the interaction depth \Z), are next applied to the combined synthetic AC events. 
Interactions on the TPC electrodes and in the xenon gas above the liquid surface contribute significantly to the isolated-S2 event rate, motivating a selection in a high-dimensional feature space as in~\cite{xenon1t_s2only}.
In this analysis, a gradient boosted decision tree (\GBDT)~\cite{gbdt} ensemble is trained using the scikit-learn package~\cite{scikit-learn} to optimize the signal and AC background discrimination based on the S2 area, the S2 rise time, the fraction of S2 area on the top array of PMTs, and \Z.
The \GBDT selection reduces the AC background by $\SI{70}{\percent}$ while accepting $\geq\SI{85}{\percent}$ of \beight~\cevns events.

A background control region with S2 $<\SI{120}{\PE}$ contains $>\SI{50}{\percent}$ of the AC background, and is excluded from the search for \beight~\cevns due to its low detection probability. 
After closer inspection of the candidate waveforms in the control region, four events whose S1s contain more than one hit in the same channel, possibly due to after-pulsing of the PMTs~\cite{xenon1t_pmt}, were removed.
Twenty-three events remain, consistent with the AC background prediction of $27.7\pm1.4$ events in the control region. 
Though the methods above yield a $\leq\SI{5}{\percent}$ uncertainty on the AC background, we conservatively use an uncertainty of $\SI{20}{\percent}$ in the analysis to reflect the statistical uncertainty from the control region, but find that the \cevns search is not strongly dependent on the uncertainty value within this range. \fref{fig:datascatter} shows the AC model, events failing the GBDT cut, and science data projected onto \Z~and quantiles of~\prevstwodt.

Neutrons originating from radio-impurities inside detector materials produce NRs in the TPC, but the tight ROI reduces these to $0.039^{+0.002}_{-0.004}$ events.
To limit the electronic recoil (ER) background dominated by $\beta$ decays of $^{214}$Pb (a daughter of $^{222}$Rn), we additionally require $\mathrm{c}S2_b$, the S2 area in the bottom array after a position-dependent correction~\cite{xenon1t_sr1}, to be $<250$~PE.
This reduces the ER background to $0.21\pm0.08$ events in the ROI, leading to a \SI{4.2}{\percent} absolute acceptance loss for \cevns. 
The same simulation procedure described in \cite{xenon1t_analysis2}~is used to assess the neutron and ER backgrounds, as well as the associated uncertainties.
The selection on $\mathrm{c}S2_b$ has negligible effect on the AC background.

\newcolumntype{b}{X}
\newcolumntype{s}{>{\hsize=.5\hsize}X}
\newcolumntype{k}{>{\centering\arraybackslash}s}
\begin{table*}[ht]
\centering
\begin{tabularx}{1.8\columnwidth}{b | s | k k k k k |  k k}
\hline \hline
\multicolumn{2}{c|}{S1 hit properties} & \multicolumn{5}{c|}{Science Data}& \multicolumn{2}{c}{AC validation region}\\
\hline
Hit Category &LHA& AC & ER & Total BG & \cevns & Data & Expected & Data\\
\hline\hline
\multirow{2}{*}{2 Hits, 1+ in TA} & $\geq2\mathrm{PE}$ & 0.09 & 0.01 & 0.10 & 0.13 & 0 & 0.25 & 0 \\
 & $<2\mathrm{PE}$ & 3.54 & 0.04 & 3.58 & 0.44 & 4 & 9.45 & 10 \\
\multirow{2}{*}{2 Hits, 0 in TA} & $\geq2\mathrm{PE}$ & 0.03 & 0.03 & 0.06 & 0.23 & 0 & 0.11 & 0 \\
 & $<2\mathrm{PE}$ & 1.47 & 0.09 & 1.58 & 0.79 & 2 & 4.07 & 4 \\
\multirow{2}{*}{$3$ Hits} & $\geq2\mathrm{PE}$ & 0.00 & 0.01 & 0.02 & 0.17 & 0 & 0.03 & 0 \\
 & $<2\mathrm{PE}$ & 0.01 & 0.03 & 0.05 & 0.36 & 0 & 0.09 & 0 \\
\hline\hline
\multicolumn{2}{c|}{Total} & 5.14 & 0.21 & 5.38 & 2.11 & 6 & 14.00 & 14 \\
\hline\hline

    \end{tabularx}
    \caption{Signal and background expectation values and observed event counts in six S1 hit classes based on number of S1 PMT hits in total, the number in the top array (TA), and the largest hit-area (LHA).
    Expectation values are computed for the nominal (NEST best fit) \qy, \ly, and \beight neutrino flux for the \SI{0.6}{\tonneyears} exposure. The neutron background is not shown separately in the table as it is significantly smaller than AC and ER, but is included in the background total.
    The last two columns show the result from the AC validation region, where the expectation value is dominated (\SI{97}{\percent}) by AC events, with the remainder from the expected \beight~\cevns leakage.
    The relative uncertainties on the background and signal expectations are described in the text.
    }
    \label{tab:muincategories}
\end{table*}

In the interpretation of the data, we utilize several features that differ between true S1-S2 events and AC.
Lone hits are spread uniformly across the top and bottom PMT arrays, whereas scintillation light from the LXe volume mostly falls on the bottom array.
Furthermore, 
an S1 with more than \SI{2}{\PE} on one PMT is very unlikely to be part of an AC, since most lone hits in XENON1T consist of a single photoelectron. 
We split the data into six ``hit categories" according to the number and arrangement of S1 hits, and the largest hit-area (LHA), listed in \tref{tab:muincategories}.

\itsec{Inference}
We analyze the data with a statistical model adapted from~\cite{xenon1t_analysis2}, with three continuous analysis dimensions; \Stwo, $Z$, and the quantiles of equal signal acceptance in \prevstwodt. The likelihood for XENON1T is the product of the likelihoods for each hit category, indexed with $i$: 
\begin{equation}
\begin{aligned}\large
    \centering
    \lagr_\xesub(\F,\qy,\ly,\nuiss) =&  \prod_{i=1}^6 \lagr_{i}(\F, \qy,\ly,\nuiss)\times\\
    &\prod_m [\lagr_m(\nuis_m)].
    \label{eqn:ll_total}
\end{aligned}
\end{equation}
Here, 
$\nuiss$ are the nuisance parameters. 
The extended unbinned likelihood terms $\lagr_i(\F, \qy,\ly,\nuiss)$ are of the same form as Eq.~(20) in ~\cite{xenon1t_analysis2}, and include models in \Stwo, $Z$ and \prevstwodt~for the \beight~\cevns~signal and AC, ER, and neutron backgrounds. 
The background component rates $\nuis_m$ are constrained by the external measurement terms $\lagr_m(\nuis_m)$.

For the~\beight~\cevns search, the nuisance parameters are the expectation values of the backgrounds, each with a constraint term, as well as the NR response parameters \qy and \ly.
The total likelihood used in the \cevns search is the product of $\lagr_\xesub$, defined in \eref{eqn:ll_total}, and external constraints on \qy and \ly, as detailed above. 
For these results, the models of \cevns, DM, and the neutron background change both in shape and expectation value with \qy and \ly. 
The \cevns discovery significance as well as DM upper limits are computed using the log-likelihood-ratio test statistic 
calibrated with toy Monte-Carlo (toy-MC) simulations~\cite{PDG,xenon1t_analysis2}.

To construct confidence intervals in \F, \qy, and \ly, we define a test statistic from the sum of profiled log-likelihoods of XENON1T and external constraints. 
By including external measurements of \qy, we can constrain \ly. 
Since the \cevns signal spans a narrow energy range, we use a constant \ly value to construct the intervals. This allows us to make use of the degeneracy between \F\ and the NR response parameters \qy and \ly, all three of which primarily affect the \cevns expectation value. 
Details on the construction of  these confidence intervals may be found in the Supplemental Material.

By including external constraints on \F, \qy, and \ly, this analysis can be used to consider physics processes beyond the Standard Model. 
We consider a benchmark model in which non-standard neutrino interactions modify the \cevns cross section~\cite{COHERENTdiscovery,COHERENT,NSI_theory}. Our confidence interval on \F\ assuming the Standard Model cross section can be reinterpreted as a confidence interval on the modified \cevns~cross section if we use the externally measured value of \F.
We also consider DM-nucleus interactions, including \cevns as a background contribution, and \qy and \ly as nuisance parameters. 
We use the same profile construction approach to compute upper limits as~\cite{xenon1t_analysis2}, including a power constraint~\cite{PCL}. 

\begin{figure}[t]\begin{center}
\includegraphics[width=0.95\columnwidth]{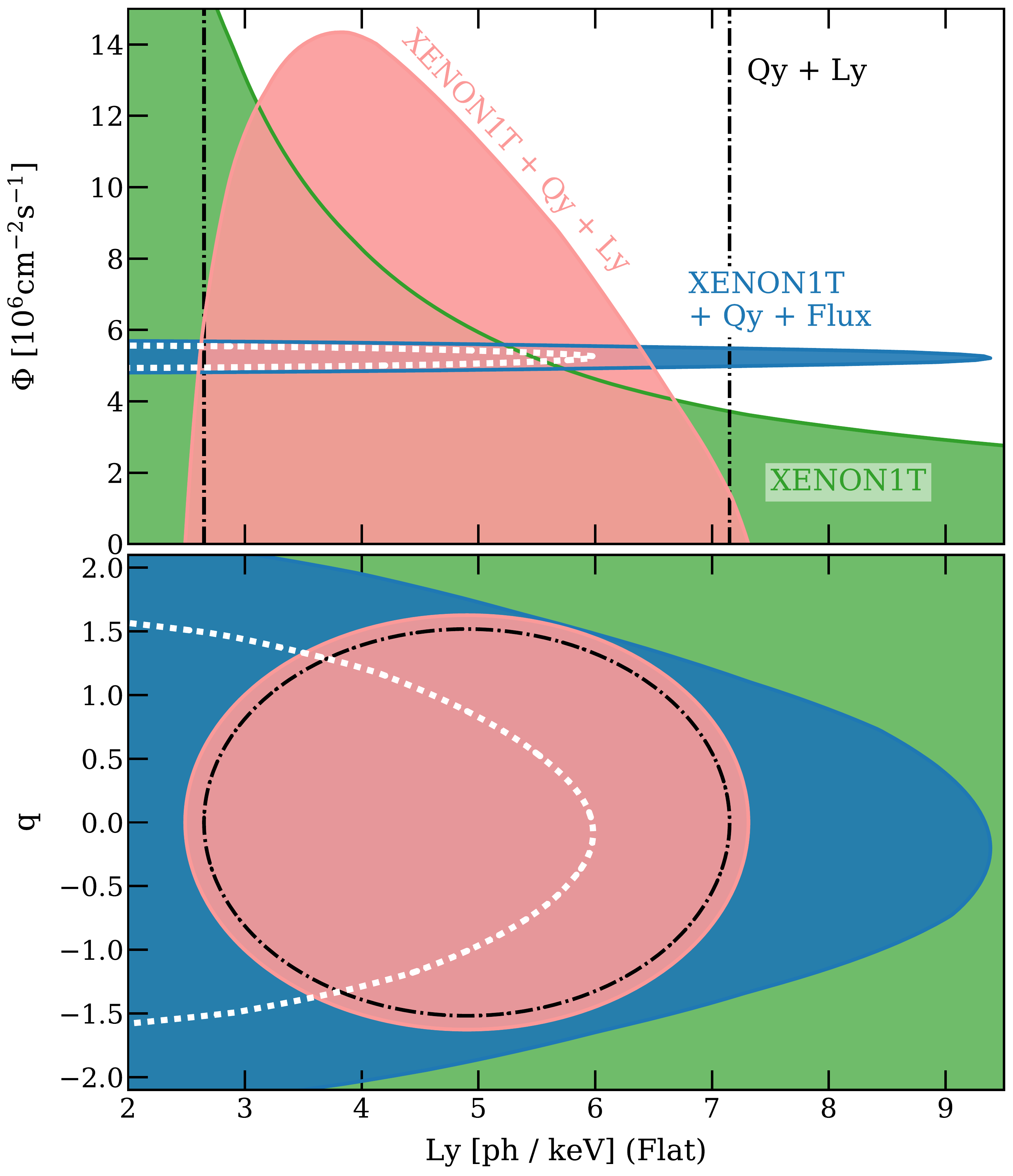}
\caption{Projections of the \SI{90}{\percent} confidence volumes in \ly and \F\ (top), and in \ly and the \qy interpolation parameter $q$ (bottom). The green area shows constraints using only the XENON1T data. Combining the XENON1T data and external constraints on \qy~\cite{lenardo} and \ly~\cite{lux2016lowenergy,dongqing_thesis} (shown in black dash-dotted lines) 
gives the confidence interval shown in pink, and an upper limit on \F. Conversely, combining the XENON1T data and constraints on \F~\cite{SNO}\ and \qy yields the dark blue interval and upper limits on \ly. The dashed white line displays the \SI{68}{\percent}~confidence interval. \ly is assumed constant in the \beight~\cevns~ROI for these constraints.}\label{fig:blob_interval}

\end{center}\end{figure}

\itsec{Results}
We estimated the probability of observing a $3\sigma~(2\sigma)$ \cevns excess in this data to be $\SI{20}{\percent}~(\SI{50}{\percent})$ for the nominal (NEST) values of \qy, \ly. 
Inverting the \GBDT cut gave an AC-rich validation region that was unblinded first (\tref{tab:muincategories}). Background-only goodness-of-fit (GOF) tests using a binned Poisson likelihood were performed on the validation region, both for the six S1 hit categories and in the continuous analysis space, with p-values of $0.95$ and $0.33$, respectively, which exceeded the $0.05$ validation criterion.
The science dataset was unblinded following the successful validation region unblinding. Six events were found, as listed in \tref{tab:muincategories}. The events are compatible with the background-only hypothesis, with a \cevns discovery significance of $p>0.50$. The same GOF tests used to assess the validation region unblinding show good agreement, with $p=0.64$ and $p=0.72$, respectively. 
The XENON1T confidence interval in \F, \qy, and \ly does not strongly constrain any of the parameters due to the significant correlation in particular between \F\ and \ly, as shown by the green shaded region in \fref{fig:blob_interval}~(top). 
On the other hand, \F\ can be constrained if the external constraints on \qy and \ly are included, as shown in the pink region, with a \SI{90}{\percent} upper limit on \F\ of \SI{1.4e7}{\cm^{-2}\second^{-1}}. 
The blue region in \fref{fig:blob_interval}~shows the confidence interval from a combination of the XENON1T likelihood, constraints on \F~\cite{SNO}, and on \qy. The \SI{90}{\percent} upper limit on \ly (assumed constant over the $0.9-1.9$~\kev~energy range) is \SI{9.4}{\pperkev}.

\begin{figure}[!htb]\begin{center}
\includegraphics[width=0.9\columnwidth]{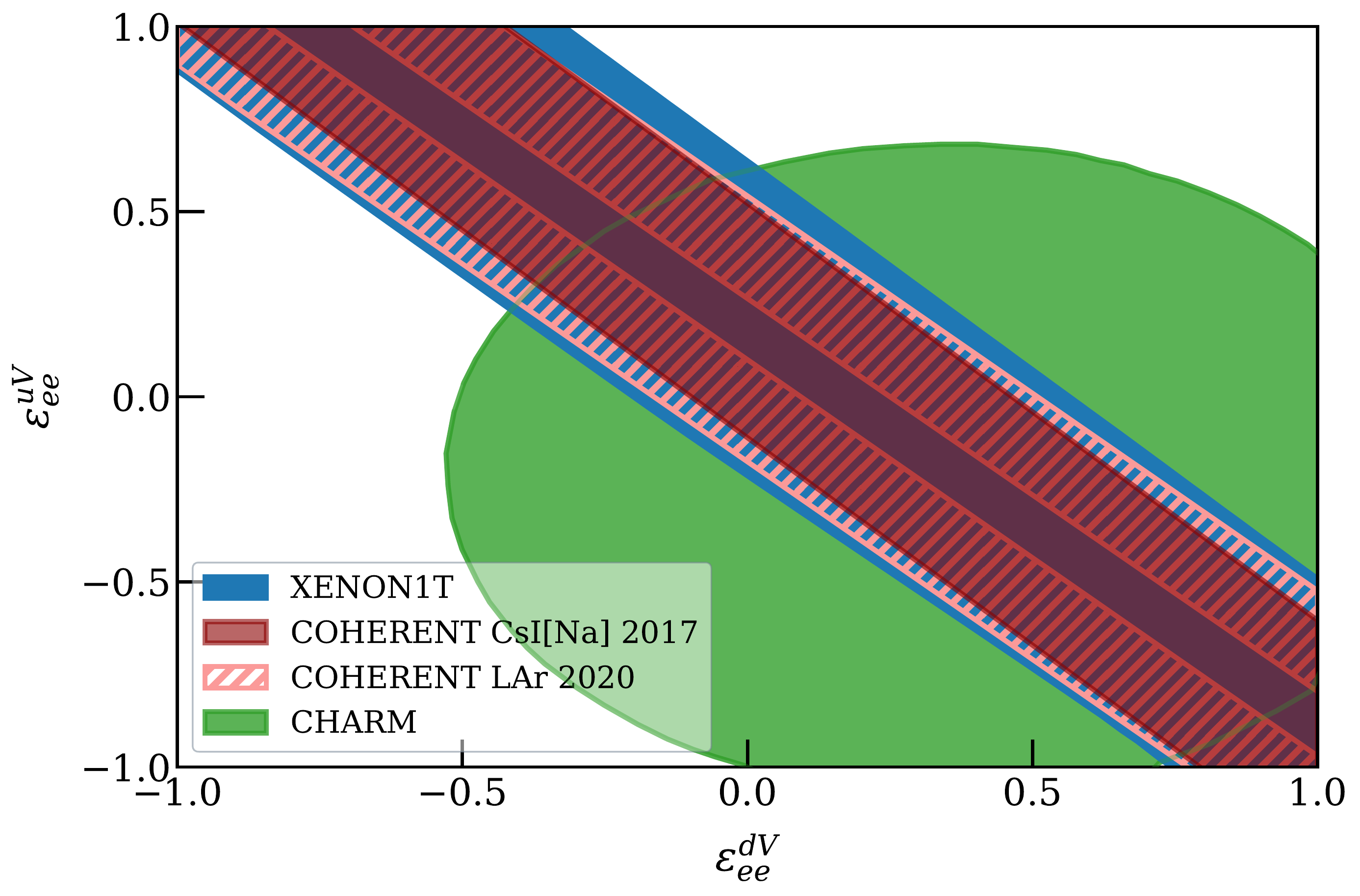}
\includegraphics[width=0.9\columnwidth]{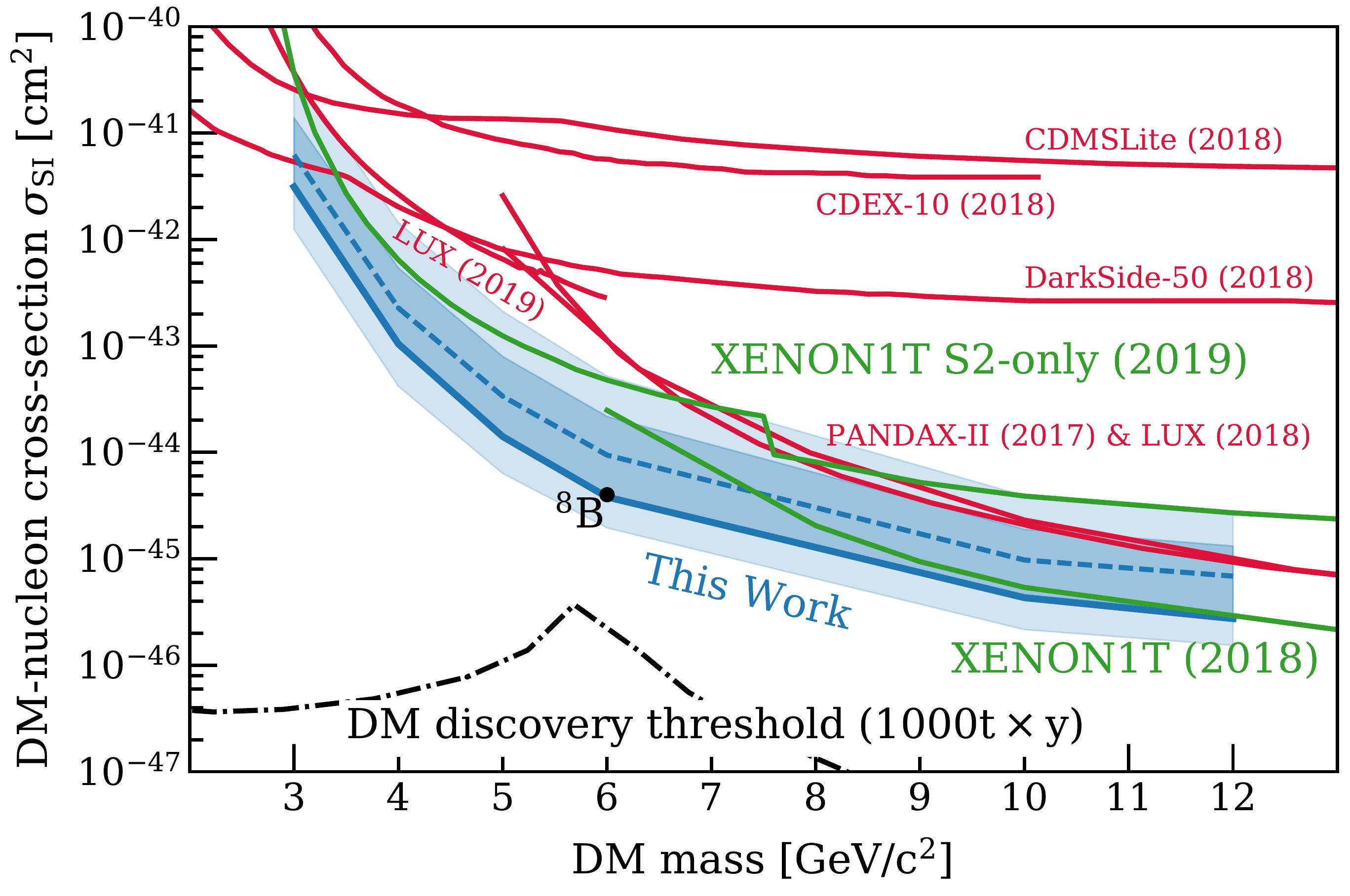}
\caption{Constraints on new physics using XENON1T data.
Top: Constraints on non-standard vector couplings between the electron neutrino and quarks, where the XENON1T \SI{90}{\percent} confidence interval (light blue region) is compared with the results from COHERENT~\cite{COHERENTdiscovery, COHERENT} (pink and dark red regions) and CHARM~\cite{CHARM} (green).
Bottom: The \SI{90}{\percent} upper limit (blue line) on the spin-independent DM-nucleon cross section $\sigma_\mathrm{SI}$ as function of DM mass. Dark and light blue areas show the $1\sigma$ and $2\sigma$ sensitivity bands, and the dashed line the median sensitivity. Green lines show other XENON1T limits on $\sigma_{SI}$ using the threefold tight-coincidence requirement~\cite{xenon1t_sr1} and an analysis using only the ionization signal~\cite{xenon1t_s2only}, and other constraints~\cite{CDMSlite, cdex, darkside, lux_2017, pandax_2017, lux_singlephoton} are shown in red. 
The dash-dotted line shows where the probability of a $3\sigma$ DM discovery is \SI{90}{\percent} for an idealized, extremely low-threshold (\SI{3}{\eV}) xenon detector with a \SI{1000}{\tonneyears} exposure~\cite{nufloor_2014}. The black dot denotes DM that has a recoil spectrum and rate identical to the \beight~neutrinos.
}\label{fig:new_physics}
\end{center}\end{figure}

In the benchmark model of non-standard neutrino interactions considered, the electron neutrino has vector couplings to the up (u) and down (d) quarks of $\varepsilon_{ee}^{dV}$ and $\varepsilon_{ee}^{uV}$, respectively~\cite{COHERENTdiscovery,COHERENT,NSI_theory}.
The \SI{90}{\percent} confidence interval for $\varepsilon_{ee}^{dV}$ and $\varepsilon_{ee}^{uV}$ from XENON1T data is shown in light blue in \fref{fig:new_physics}~(top). 

The result for a spin-independent DM-nucleus interaction is shown in \fref{fig:new_physics}~(bottom).
This constraint improves on previous world-leading limits~\cite{xenon1t_sr1,xenon1t_s2only} in the mass range between $\SI{3}{\gevm}$ and $\SI{11}{\gevm}$ by as much as an order of magnitude. The limit lies at roughly the $15$th percentile, reflecting the downwards fluctuation with respect to the background model (including \cevns), but is not extreme enough to be power-constrained.

\itsec{Outlook}
The XENONnT experiment, currently being commissioned at LNGS, aims to acquire a \SI{20}{\tonneyears} exposure~\cite{xenonnt_sensitivity}.
As the isolated-S1 rate scales up with the larger number of PMTs and the isolated-S2 rate with the detector surface area, the AC background will be the biggest challenge for the discovery of \beight\,\cevns.
The AC background modeling and discrimination techniques used in this analysis will improve the sensitivity of XENONnT to \beight~\cevns and low-mass DM. 
The novel cryogenic liquid circulation system developed to ensure efficient purification in XENONnT will mitigate the reduction of S2s due to impurities, improving the acceptance of low-energy NRs from \beight neutrinos and DM. 
Additionally, the data will be analyzed in a triggerless mode to minimize efficiency loss and better understand the AC background. Together with the significantly larger exposure, these techniques give XENONnT strong potential to discover \beight~\cevns.

The large uncertainty in both \qy and \ly will be the dominant systematic in constraining new physics from DM and non-standard neutrino interactions. 
Improving these uncertainties by calibrating NRs in LXe using \textit{in situ} low energy neutron sources~\cite{ybe} 
and dedicated detectors~\cite{lenardo} can crucially improve the sensitivity of next-generation experiments to both \beight~\cevns and light DM.

\itsec{Acknowledgements}
We would like to thank Matthew Szydagis and Ekaterina Kozlova for useful discussions concerning the NEST model.
We gratefully acknowledge support from the National Science Foundation, Swiss National Science Foundation, German Ministry for Education and Research, Max Planck Gesellschaft, Deutsche Forschungsgemeinschaft, Helmholtz Association, Netherlands Organisation for Scientific Research (NWO), Weizmann Institute of Science, ISF, Fundacao para a Ciencia e a Tecnologia, Région des Pays de la Loire, Knut and Alice Wallenberg Foundation, Kavli Foundation, JSPS Kakenhi in Japan and Istituto Nazionale di Fisica Nucleare. This project has received funding or support from the European Union’s Horizon 2020 research and innovation programme under the Marie Sklodowska-Curie Grant Agreements No. 690575 and No. 674896, respectively. Data processing is performed using infrastructures from the Open Science Grid, the European Grid Initiative, and the Dutch national e-infrastructure with the support of SURF Cooperative. We are grateful to Laboratori Nazionali del Gran Sasso for hosting and supporting the XENON project.

\bibliography{bibliography}

\ifnum \value{includeappendix}>0 {\newpage 

\section{Supplemental Material}
\setcounter{page}{1}


\subsection{Waveform Simulation}
The S1 and S2 detection efficiencies in the \cevns ROI cannot easily be measured with a calibration source. 
Therefore, we use a waveform simulation, which produces PMT waveforms in the \cevns ROI, to calculate those efficiencies. 

Some of the S1s detected by two or more PMTs do not meet the requirement that hits on those PMTs occur within \SI{50}{\ns}. 
The fraction of the S1s passing this tight-coincidence requirement thus correlates with the S1 width. 
We use an exponential function to describe the distribution of photons detected by PMTs in the simulation to facilitate tuning of the S1 width. 
The S1 time distribution is independent of the number of hits in XENON1T data. 
This allows us to calibrate the exponential function by matching simulated S1s to those in data.

Four more detector effects are included in the simulation: the probability that the PMT photocathode emits two photoelectrons when absorbing one photon, the electronic noise level, the single photoelectron spectrum of the PMTs, and PMT after-pulses. 
The full simulation process establishes the relation between the number of detected photons and the size of the S1 and S2 \cite{xenon1t_analysis2}.

The mean and spread of the S1 width distribution vary with the size of the S1.
Simulated waveforms and XENON1T data are processed with the same software.
The S1 width parameter in the simulation is tuned to minimize the chi-square between simulated and observed mean width as shown in \fref{fig:wfpmatch}. 

The software trigger efficiency of the S2 varies with its size and the position of the event. Events from the deeper part of the detector produce wider S2s, and have a lower trigger efficiency.  
Specifically, in waveform simulation, we use effective models to reproduce the diffusion, size, and temporal distribution of ionization signals. Together with the four detector effects mentioned above, the simulation output is compared to background S2s originating on the detector wall in both width and triggered fraction, since wall events have a smaller S2 size due to charge loss on surfaces. The excellent matching between simulated and wall events, shown in \fref{fig:wfpmatch} and \fref{fig:threshold},  validates the response of the detector to small S2s.

\begin{figure*}[ht]
    \centering
    \includegraphics[width=2.\columnwidth]{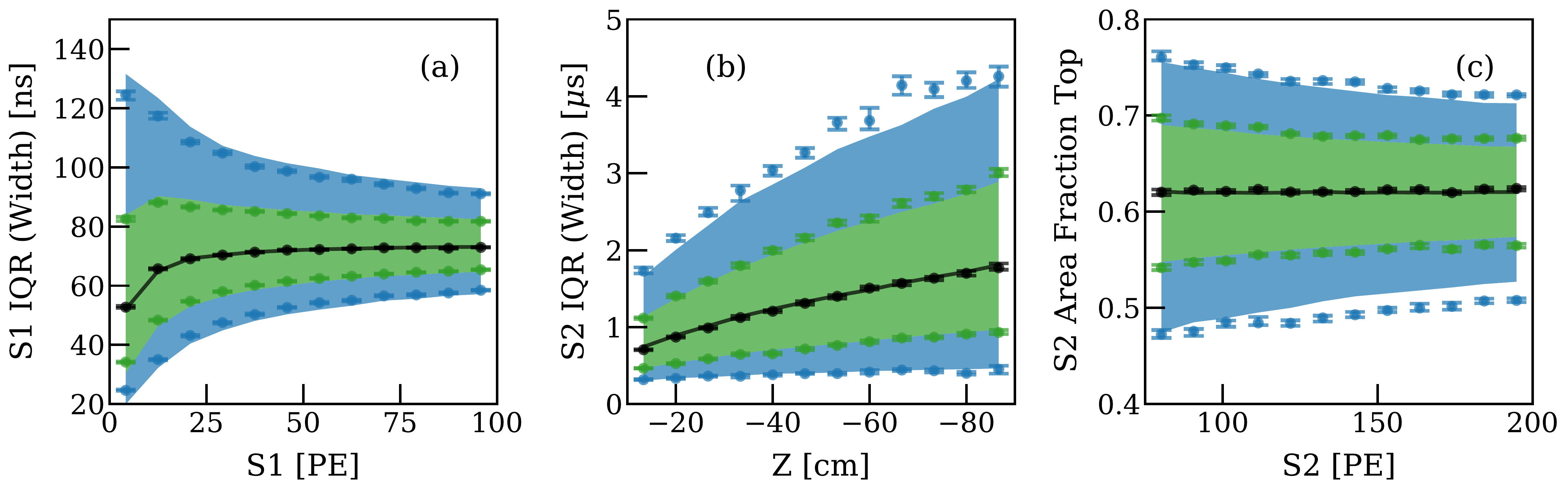}
    \caption{Matching of the S1 and S2 properties of the waveform simulation and detector data originating from the TPC wall. The plots show (a) the interquartile range (IQR), the range of time covering the central \SI{50}\percent~area, as a function of S1 size, (b) IQR of $\Stwo<\SI{200}{\PE}$ as a function of depth (\Z), and (c) the fraction of signal detected by the top PMT array in each  S2 ($\SI{-10}{\cm}< Z < \SI{90}{\cm}$). The dots denote quantiles of the detector data, corresponding to $\pm$ 2 $\sigma$ (blue), $\pm$ 1 $\sigma$ (green), and median (black). Colored bands show the same quantiles with the simulation data. Both detector and simulation data are events close to the wall, with the same position, S1 size, and S2 size distributions. 
}
    \label{fig:wfpmatch}
\end{figure*}

\begin{figure*}[ht]
    \centering
    \includegraphics[width=2.0\columnwidth]{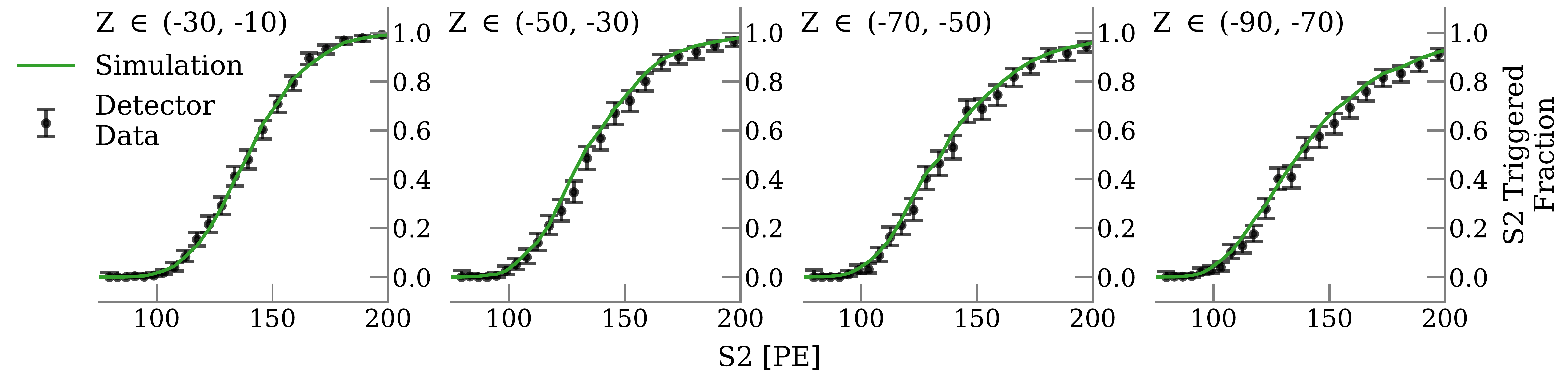}
    \caption{Efficiencies (fraction of S2 passed) of the software trigger as a function of the S2 size at different depth (\Z) ranges (in cm) show agreement between the simulation and detector data. The black dots correspond to the events in detector data, that were triggered by their S1. The green line shows the fraction of simulated S2 passing the trigger, produced with the same position distribution as the S1-triggered detector data.
}
    \label{fig:threshold}
\end{figure*}

\subsection{Signal expectation}
From the standard solar model, the energy of solar~\beight neutrinos is below $\sim\SI{20}{\mev}$, giving a maximum momentum transfer $q_{\mathrm{max}} \sim\SI{40}{\mev}$, much smaller than the Z boson mass~\cite{billard}. 
Under this condition, the Standard Model predicts that the tree-level differential \cevns cross section is given by:
\begin{equation}
    \frac{\mathrm{d}\sigmac}{\mathrm{d}E_r}=\frac{G_F^2}{4\pi}Q_w^2M\left(1-\frac{ME_r}{2E_\nu^2}\right)F(E_r)^2,
\end{equation}
where $E_r$ is the NR energy, $G_F$ is the Fermi constant, $M$ is the target nuclear mass of the recoiling atom, $E_\nu$ is the incoming neutrino energy, $F(E_r)$ is the nuclear form factor, and $Q_w$ is the nuclear weak charge~\cite{cevnsproposal}. Here, we have neglected the contribution from the hadronic axial-vector current, because the spin-dependent structure factors are negligible compared with spin-independent structure factors for xenon \cite{xenon1t_SD}. 
Since $M\gg E_\nu\gg E_r$, terms of higher order in $E_r/E_\nu$ are dropped as well. 

We also consider a non-standard interaction following \cite{COHERENT, NSI_theory}, where the weak charge in electron neutrino scattering is replaced by $Q_w\to \widetilde{Q}_w= N(1+2\varepsilon_{ee}^{uV}+4\varepsilon_{ee}^{dV}) + Z(4\sin^2\theta_w-1+4\varepsilon_{ee}^{uV}+2\varepsilon_{ee}^{dV})$, with two non-standard couplings $\varepsilon_{ee}^{uV}$ and $\varepsilon_{ee}^{dV}$. Neutrino oscillation must be included, since our model assumes that only electron neutrinos have non-zero non-standard interactions. 
In the energy range of solar~\beight neutrinos, their oscillation to other flavors through interactions with matter in the Sun (the MSW effect) is important \cite{MSW}. In the standard model, this effect can be interpreted as an equivalent index of refraction $n=1+\sqrt{2}G_FN_e/E_\nu$ for electron neutrinos, with $N_e$ being the electron number density. Our model assumes two additional non-standard interactions $\varepsilon_{ee}^{uV}$ and $\varepsilon_{ee}^{dV}$, so the index of refraction should be modified to be $\widetilde{n}=1+\sqrt{2}G_F(N_e+\varepsilon_{ee}^{uV}N_u+\varepsilon_{ee}^{dV}N_d)/E_\nu$, where $N_u$ ($N_d$) is the number density of up (down) quarks. Thus, the inclusion of non-standard interactions also makes the survival probability of electron neutrinos $P_e$ epsilon-dependent \cite{MSW_NSI}. The neutrino oscillation parameters in the following calculation are from \cite{neutrino_params}. Using $\Phi=\SI{5.25 \pm 0.2e6}{ cm^{-2} s^{-1}}$ and letting $\eta(E_r)$ be the NR acceptance, the final expected \cevns rate from solar \beight neutrinos is
\begin{equation}
    R(\varepsilon_{ee}^{uV}, \varepsilon_{ee}^{dV})=\frac{1}{M}\int\frac{\mathrm{d}\sigma}{\mathrm{d}E_r}\frac{\mathrm{d}\Phi}{\mathrm{d}E_\nu}\eta(E_r)\,\mathrm{d}E_\nu\mathrm{d}E_r,
\end{equation}
where $\mathrm{d}\sigma/\mathrm{d}E_r$ is given by
\begin{equation}
    \frac{\mathrm{d}\sigma}{\mathrm{d}E_r}=\frac{\mathrm{d}\sigmac}{\mathrm{d}E_r}\frac{\widetilde{Q}_w^2}{Q_w^2}P_e+\frac{\mathrm{d}\sigmac}{\mathrm{d}E_r}(1-P_e).
\end{equation}
So the upper limit on $\Phi$ can be converted into $\varepsilon_{ee}^{uV}$-$\varepsilon_{ee}^{dV}$ space by solving:
\begin{equation}
    \left\langle R(\varepsilon_{ee}^{uV}, \varepsilon_{ee}^{dV})\right\rangle<\frac{\Phi_\mathrm{limit}}{\Phi}\left\langle R(\varepsilon_{ee}^{uV}=0, \varepsilon_{ee}^{dV}=0)\right\rangle,
\end{equation}
where $\langle\cdot\rangle$ denotes the isotopic average (assuming natural abundances in xenon), and $\Phi_\mathrm{limit}=\SI{1.4e7}{\cm^{-2}\second^{-1}}$ is the upper limit on $\Phi$ (see Results section).

\subsection{More details on the AC background}
The rates of isolated S1s and isolated S2s are significantly increased following high-energy events, mainly due to gamma-ray backgrounds. In XENON1T, we found that the rate of single-electron S2s and lone hits on PMTs are correlated with~\prevstwodt. \fref{fig:pres2vdt} shows the distribution of~\prevstwodt for both isolated S1s and high-energy events themselves. The distribution for high-energy events reflects that of signal events, since neither are correlated with preceding S2s. Thus, a selection requiring \prevstwodt$\leq12\,$PE/$\mu$s rejects 65\% of isolated S1s (and consequently AC events) with 87\% signal acceptance.

\begin{figure}[t]\begin{center}
    \includegraphics[width=0.9\columnwidth]{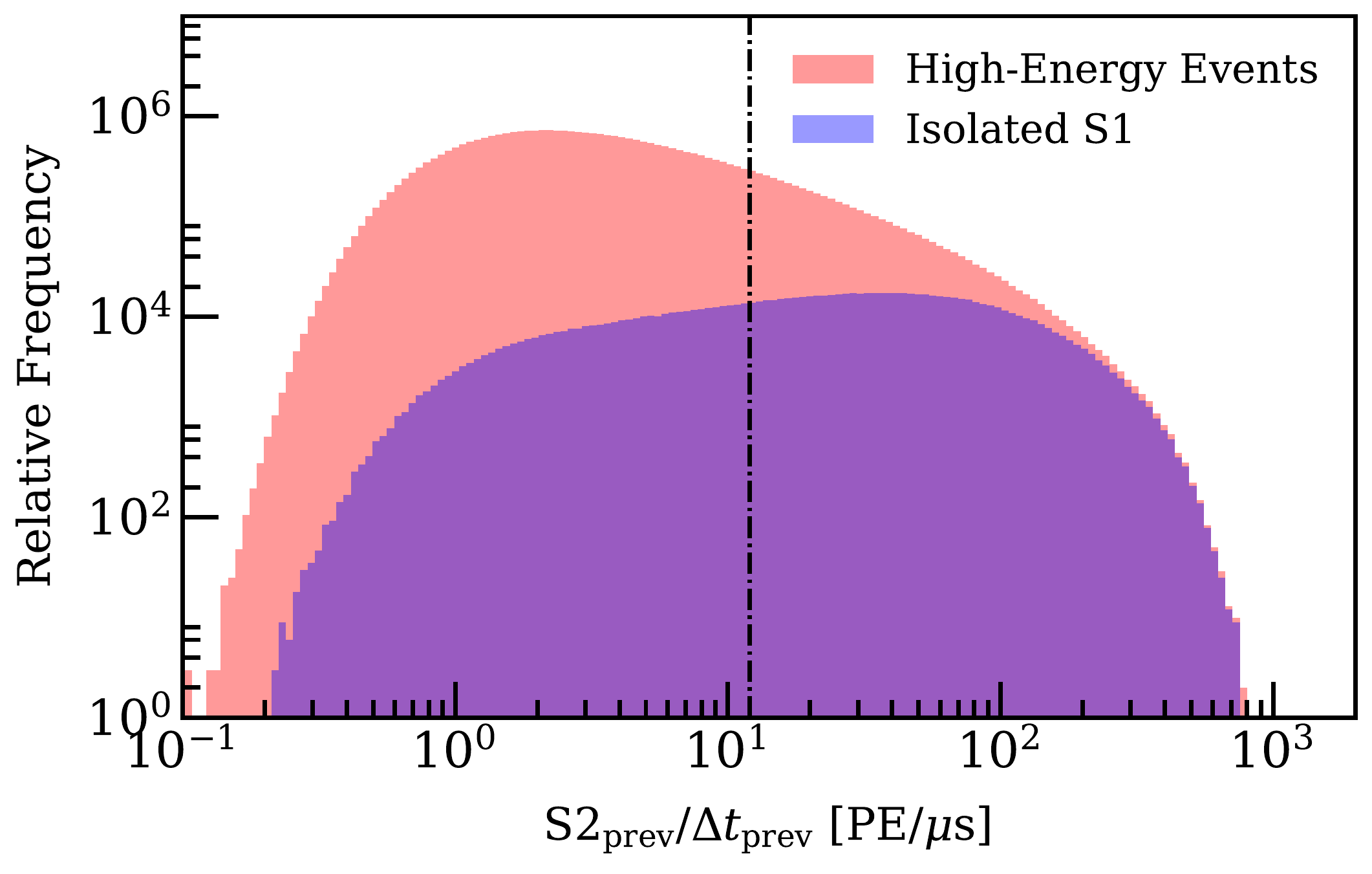}
    \caption{
    Distribution in \prevstwodt\, space of both high-energy events and isolated S1s immediately preceding them. At large \prevstwodt\, values, a significant fraction of high energy events contain an isolated S1 before the trigger. The discrimination between signal events (which have the same distribution as high-energy events) and isolated S1s allows the selection \prevstwodt$\leq12$PE/$\mu$s, shown as a vertical black dash-dotted line, which rejects 65\% of background with 87\% signal acceptance.
}
    \label{fig:pres2vdt}
\end{center}\end{figure}

Although the selection on \prevstwodt\ also suppresses the rate of isolated S2s, those that remain near the 80\,PE threshold are still correlated with \prevstwodt. To remove this correlation, we utilize the horizontal (X,Y) positions of isolated S2s, calculated from fitting their PMT distribution patterns, similar to \cite{xenon1t_analysis1}. We investigate the horizontal spatial distance of isolated S2s from previous high energy events, $\sqrt{(\mathrm{X}-\mathrm{X}_{\mathrm{prev}})^2+(\mathrm{Y}-\mathrm{Y}_{\mathrm{prev}})^2}$, to quantify the correlation between them, as shown in \fref{fig:pres2vdx}. A distinctive population with small $\sqrt{(\mathrm{X}-\mathrm{X}_{\mathrm{prev}})^2+(\mathrm{Y}-\mathrm{Y}_{\mathrm{prev}})^2}$ values is seen near the 80\,PE threshold. A cut, shown as a red line, rejects $>99$\% of events that occur at the same (X, Y) position as the preceding event but are reconstructed with non-zero mean-squared distance due to uncertainties in the reconstructed positions. The acceptance of this cut as a function of S2 is studied by randomly associating two uncorrelated events, and ranges from 92\% at 120\,PE to $>99$\% for S2 $>500$\,PE. 

\begin{figure}[t]\begin{center}
    \includegraphics[width=0.95\columnwidth]{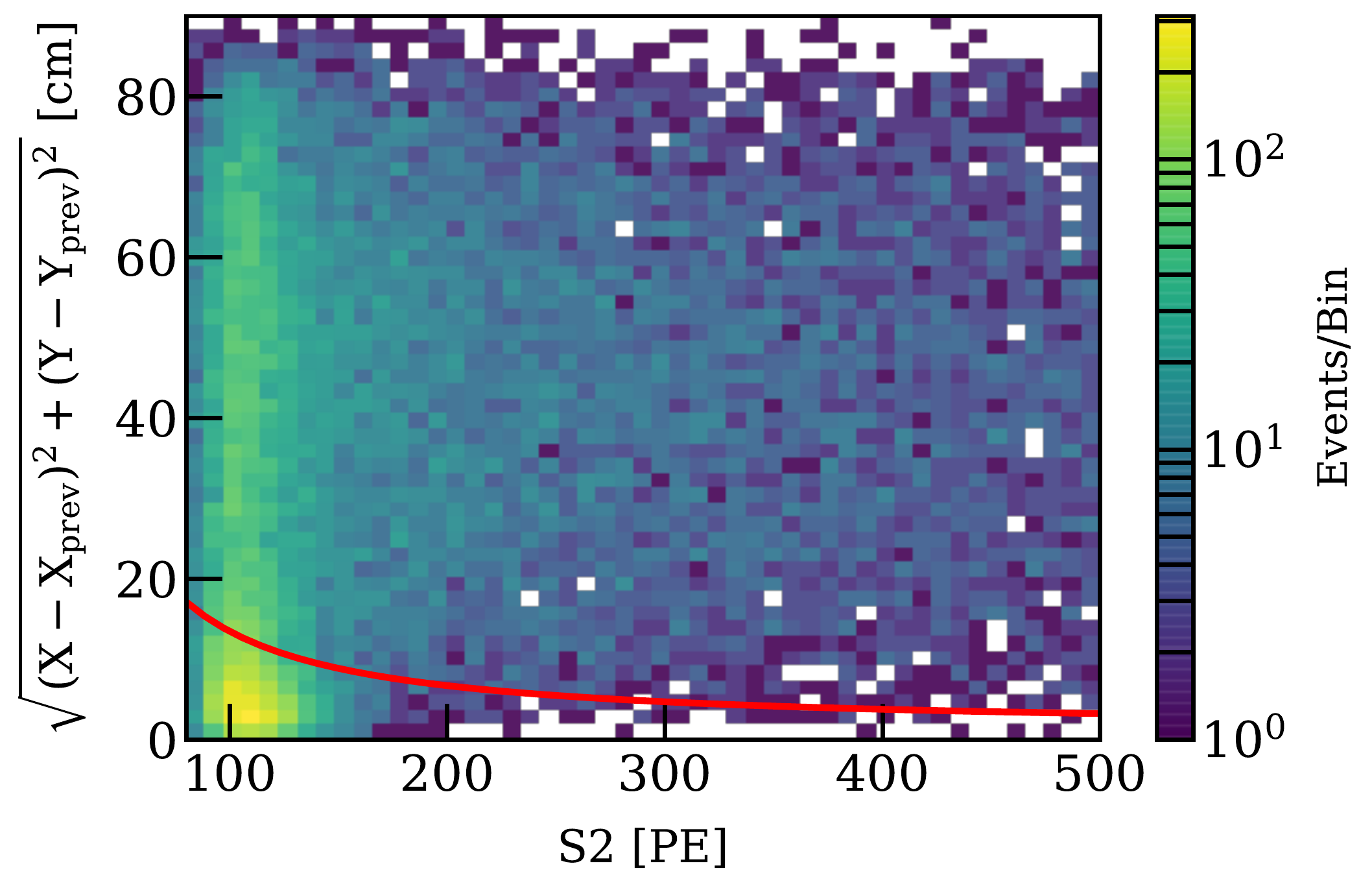}
    \caption{Horizontal spatial distance of isolated S2s with respect to previous high-energy events. Events below the red line are removed to ensure no correlation between isolated S2s and the preceding high-energy event.}
    \label{fig:pres2vdx}
\end{center}\end{figure}

\subsection{Details on constructing confidence volumes}
Since the NR response uncertainty is large, the test statistic distribution for confidence intervals will depend on the true values of \F, \qy, and \ly. 
To compute a unified confidence interval in all these parameters in the manner of~\cite{xenon1t_analysis2} and the DM results in this paper, we would have to estimate the distribution of the test statistic using toy-MC computations in these three dimensions. 
However, the strong degeneracy between these parameters allows us to avoid this extensive computation. 
For the relevant range for this search, and the low number of events expected, the \cevns model shape changes so little with \qy and \ly that inference results are not affected: Computing the discovery significance of toy-MC simulations either fitting these shape parameters, fixing them to their true values or shifting them by $+2\sigma$ each yielded no discernible bias, and a spread compatible with toy-MC variation only.
Therefore, only when computing confidence intervals on \F\footnote{Also used for the non-standard neutrino interaction result}, \qy, and \ly, the \cevns model shape is fixed, and these variables appear in the likelihood only via the expression for the expectation value of detected \cevns events, $\mu_\cevns(\F,\qy,\ly)$. 
Therefore, we can compute the profile likelihood ratio and toy-MC estimates of the test statistic distribution in the space of $\mu_\cevns$ alone. 
External constraints on \F, \ly, and \qy are implemented as terms $\lambda_\mathrm{F}$, $\lambda_\mathrm{\ly}$, and $\lambda_\mathrm{\qy}$, corresponding to the profiled log-likelihood-ratios for Gaussian measurements of each parameter.
We combine the XENON1T profiled log-likelihood ratio $\lambda_\xesub(\mu_\cevns(\F,\qy,\ly))$ and different combinations of external constraints into test statistics $\Lambda$: 
\begin{equation}\large
    \centering
    \begin{aligned}
    \Lambda_\mathrm{A}(\F,\qy,\ly) =& \lambda_\xesub(\mu_\cevns(\F,\qy,\ly))\\
    \Lambda_\mathrm{B}(\F,\qy,\ly) =& \lambda_\xesub(\mu_\cevns(\F,\qy,\ly)) +\\ 
    &\lambda_\mathrm{\qy}(\qy) + \lambda_\mathrm{\cevns}(\F)\\
    \Lambda_\mathrm{C}(\F,\qy,\ly) =& \lambda_\xesub(\mu_\cevns(\F,\qy,\ly)) +\\ 
    &\lambda_\mathrm{\qy}(\qy) + \lambda_\mathrm{\ly}(\ly)\mathrm{.}\\
    \label{eqn:ll_xenonandexternal}
    \end{aligned}
\end{equation}
For each $\Lambda$, the toy-MC results of $\lambda_\xesub(\mu_\cevns)$ is combined with random realizations of the other profiled likelihoods in a grid of $\F,\qy$, and $\ly$ to provide the 90th percentile of $\Lambda$ for each point in parameter space, which is compared with $\Lambda(\F,\qy,\ly)$ to construct confidence intervals.
The test statistic $\Lambda_\mathrm{A}$, shown in green in Fig.~3 in the main text, represents the confidence interval using the XENON1T data only. The strong anti-correlation between \F~and \ly~is apparent in Fig.~3~(top). 
To compute a confidence interval on \ly, we include constraints on \qy\cite{lenardo} and \F~\cite{SNO} in $\Lambda_\mathrm{B}$, shown in dark blue in Fig.~3 in the main text. Last, combining XENON1T, and constraints on \qy~\cite{lenardo} and \ly~\cite{lux2016lowenergy,dongqing_thesis} into $\Lambda_\mathrm{C}$ yields an upper limit on the \cevns interaction rate \F.

} \else {} \fi

\end{document}